\documentclass[lettersize,journal]{IEEEtran}
\usepackage{amsmath,amsfonts}
\usepackage{algorithmic}
\usepackage{algorithm}
\usepackage{array}
\usepackage[caption=false,font=normalsize,labelfont=sf,textfont=sf]{subfig}
\usepackage{textcomp}
\usepackage{stfloats}
\usepackage{svg}
\usepackage{url}
\usepackage{xcolor}
 
\usepackage{verbatim}
\usepackage{graphicx}
\usepackage{cite}
\ifCLASSINFOpdf
\else
\fi

\begin{document}
\title{Semantic Communication for the Internet of Sounds: Architecture, Design Principles, and Challenges}
	\author{\IEEEauthorblockN{Chengsi Liang\IEEEauthorrefmark{1}, Yao Sun\IEEEauthorrefmark{1}, Christo Kurisummoottil Thomas\IEEEauthorrefmark{2}, Lina Mohjazi\IEEEauthorrefmark{1}, and
	Walid Saad\IEEEauthorrefmark{2}}
	\IEEEauthorblockA{\IEEEauthorrefmark{1}James Watt School of Engineering,
	University of Glasgow, Glasgow, UK\\
	\IEEEauthorrefmark{2}Department of Electrical and Computer Engineering at Virginia Tech, Arlington, VA 22203 USA\\
	Email: Yao.Sun@glasgow.ac.uk}}


\author{\IEEEauthorblockN{Chengsi Liang, Yao Sun, Christo Kurisummoottil Thomas, Lina Mohjazi, and Walid Saad}

\thanks{Chengsi Liang, Yao Sun (corresponding author), and Lina Mohjazi are with the James Watt School of Engineering, University of Glasgow, Glasgow G12 8QQ, UK (e-mail: 2357875l@student.gla.ac.uk; \{yao.sun, lina.mohjazi\}@glasgow.ac.uk).}
\thanks{
Christo Kurisummoottil Thomas and Walid Saad are with  the Bradley Department of Electrical and Computer Engineering at Virginia Tech, Arlington, VA 22203, USA. (e-mail: \{christokt, walids\}@vt.edu).}
}
	
\maketitle
\begin{abstract}
The Internet of Sounds (IoS) combines sound sensing, processing, and transmission techniques, enabling collaboration among diverse sound devices. To achieve perceptual quality of sound synchronization in the IoS, it is necessary to precisely synchronize three critical factors: sound quality, timing, and behavior control. However, conventional bit-oriented communication, which focuses on bit reproduction, may not be able to fulfill these synchronization requirements under dynamic channel conditions. One promising approach to address the synchronization challenges of the IoS is through the use of semantic communication (SC) that can capture and leverage the logical relationships in its source data. Consequently, in this paper, we propose an IoS-centric SC framework with a transceiver design. The designed encoder extracts semantic information from diverse sources and transmits it to IoS listeners. It can also distill important semantic information to reduce transmission latency for timing synchronization. At the receiver's end, the decoder employs context- and knowledge-based reasoning techniques to reconstruct and integrate sounds, which achieves sound quality synchronization across diverse communication environments. Moreover, by periodically sharing knowledge, SC models of IoS devices can be updated to optimize their synchronization behavior. Finally, we explore several open issues on mathematical models, resource allocation, and cross-layer protocols.
\end{abstract}
\IEEEpeerreviewmaketitle

\section{Introduction}
The Internet of Sounds (IoS) represents a confluence of sound and music systems with the Internet of Things (IoT) \cite{turchet2023ios}. In an IoS, dedicated sound devices are designed and used for sensing, capturing, processing, actuating, and sharing sounds and sound-related information via acoustic signal processing and deep learning (DL) techniques.
The IoS integrates co-located or remotely connected sound-centric devices to work together efficiently and seamlessly. 
Furthermore, the IoS promises to support novel sound-based applications including smart home, smart healthcare such as sound-based therapies, and wildlife monitoring, among others. 

One significant challenge in the IoS is the need for synchronization, particularly for time-sensitive applications such as live concerts. To guarantee high-quality auditory perception for IoS listeners, these applications demand rapid capture, exchange, and synchronization of sounds from a variety of IoS senders within a short timeframe.
Synchronization in the IoS should take into account three major factors: \textit{sound quality}, \textit{timing} and \textit{behavior control}. 
Sound quality synchronization focuses on maintaining the fidelity and clarity of each sound stream and developing the auditory perception during the sound synchronization process. Meanwhile, \textit{timing} is the fact that the IoS requires a precise alignment of multiple sounds transmitted from different IoS senders to guarantee that they are played at the intended moments, despite their different geographical locations. Finally, synchronization extends to the \textit{behavior control} of IoS devices.
``Behavior" refers to IoS devices' role-specific actions in synchronization tasks, such as live concerts. Behavior control strategies define the roles of IoS devices and dictate which types of sounds they should prioritize based on their assigned roles. Moreover, these strategies should dynamically adapt to changing performance requirements and listener needs.

Thereby, there is a need for intelligent, adaptable communication paradigms that can support the IoS and overcome its synchronization challenges. In this regard, semantic communication (SC) \cite{chaccour2024less} could be a viable approach. Unlike traditional bit-oriented communication, SC enables a network to convey the meanings embedded within messages. SC leverages a semantic encoder and decoder with background knowledge to ensure meanings are interpreted accurately, even in the presence of significant bit errors that would typically disrupt traditional communications. 

Based on these features, we therefore envision that SC can improve all three factors of IoS sound synchronization. 
First, SC can reconstruct and interpolate missing or corrupted sound segments by leveraging semantic-aware reasoning based on context and background knowledge \cite{CKT}. Even when channels suffer from significant noise and interference, SC helps preserve sound quality during synchronization by filling in the gaps through intelligent inference.  
Furthermore, SC senders can convey only ``essential'' semantic information, thereby SC receivers can reconstruct sounds promptly. This approach compensates for the transmission latency experienced by remote devices and ensures precise timing synchronization.
Finally, SC systems can acquire new knowledge from new sensing data and listeners' feedback, and, then, update their knowledge base. By sharing the new knowledge among multiple IoS devices, these IoS devices can update their models and automatically adjust their behavior. Instead of using conventional control messages, this knowledge-based control strategy allows for system-wide improvements and dynamic optimization based on user experiences. 

\begin{figure*}[!t]
    \centering
    \includegraphics[width=0.67\textwidth]{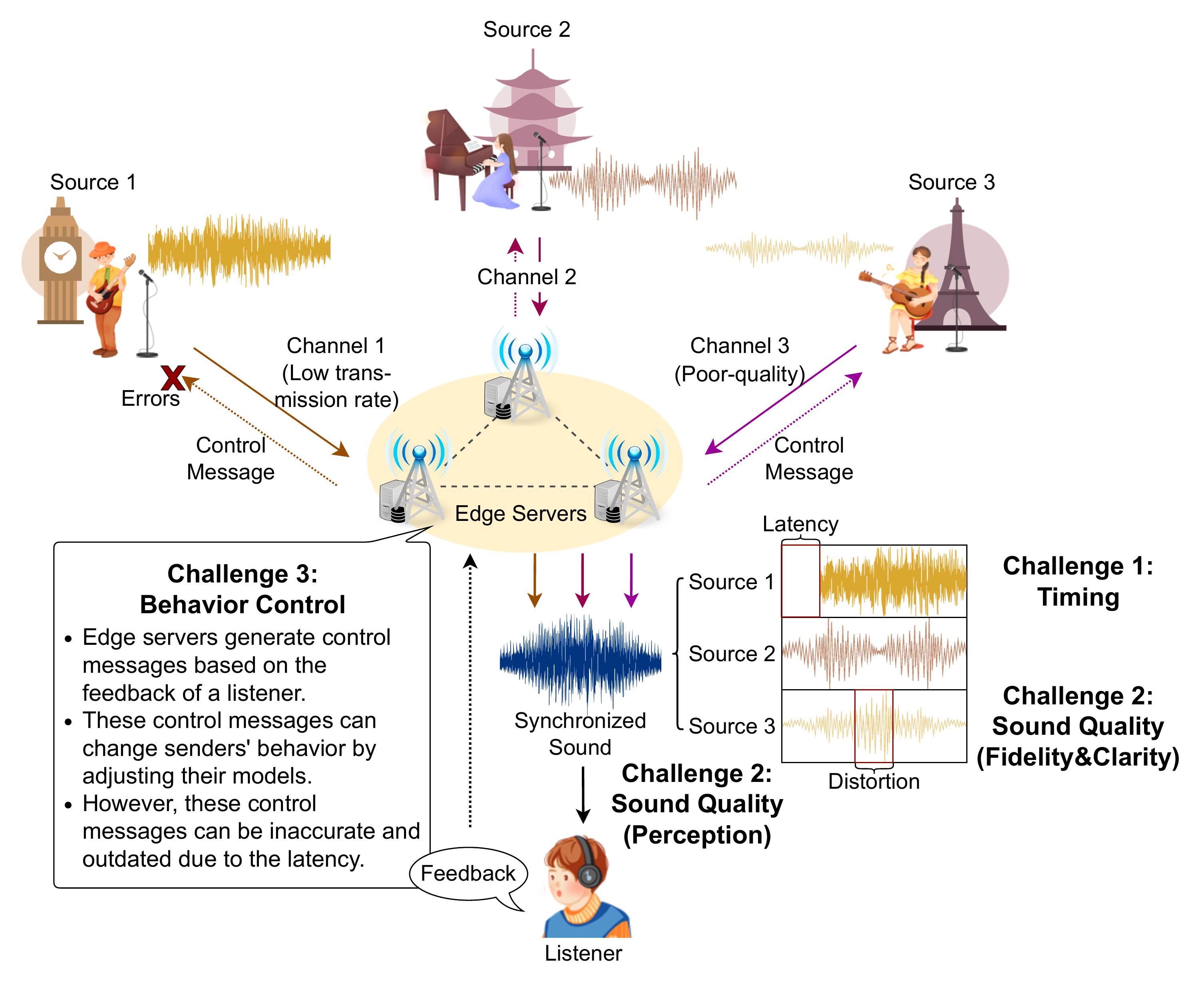}
    \caption{Three sound synchronization challenges in the IoS.}
    \label{challenge}
\end{figure*}

However, leveraging SC in the IoS requires meeting several critical challenges, that fall into three categories. \\
\textbf{Robust Semantic Encoding and Decoding:}
Unlike traditional encoder/decoder design in SC, which focuses solely on a single transmission pair, the encoder/decoder design in the IoS-centric SC should consider the synchronization of the entire network. How to design encoder/decoder to improve both the transmission quality of individual links and synchronization in a multi-user IoS is a nontrivial challenge. \\
\textbf{Latency Synchronization across Geographically Heterogeneous Devices:}
In IoS-centric SC networks, IoS listeners receive sounds from multiple geographically heterogeneous IoS senders. Since these senders may be located in different regions, the latency experienced by each device can vary significantly, which makes it challenging to achieve precise sound synchronization. 
Thus, how to implement advanced semantic-aware communication techniques is crucial for latency synchronization across geographically heterogeneous devices is another challenge.\\
\textbf{Knowledge-based Control Strategy with Semantic Effectiveness Evaluation: }
To provide users with a seamless and immersive auditory experience, IoS senders should adapt their behavior based on the listeners' feedback in real-time. In SC networks, senders can update their models periodically by learning from the knowledge (e.g., history and sensing data) without the need for control messages. However, a knowledge-based behavior control strategy is required to manage how and when to share knowledge among diverse devices in the dynamic IoS. 
Moreover, traditional communication metrics are no longer applicable for evaluating semantic effectiveness in sound synchronization. 

Recently, the authors in \cite{turchet2023ios} conducted a comprehensive survey of the IoS, exploring synchronization issues in classical communication schemes and discussing semantic-aware audio processing. However, their work did not delve into the specifics of semantic-aware communications in the IoS. Moreover, despite some recent works like \cite{tong2021federated} and \cite{weng2023deep} that investigate the use of SC for speech signals, those prior works do not extend to an IoS system. Moreover, those works do not consider cooperation among multiple users for time-sensitive sound synchronization. 

In contrast to this prior art, the main contribution of this paper is a novel SC framework that enables a robust and seamless sound synchronization in the IoS. 
We first articulate the synchronization challenges in terms of sound quality, timing, and behavior control within the IoS. We then delve into how SC can address these challenges by reasoning from context and background knowledge and facilitating knowledge sharing among IoS-related devices. Next, we propose an IoS-centric SC framework that encompasses sound devices, SC coordinators, base stations (BSs), and cloud servers. Within this framework, we introduce the transceiver design, which includes semantic encoding, semantic importance-aware transmission, and semantic decoding. Finally, we discuss open research questions for future work at the intersection of IoS and SC.

\begin{figure*}[!t]
    \centering
    \includegraphics[width=0.95\textwidth]{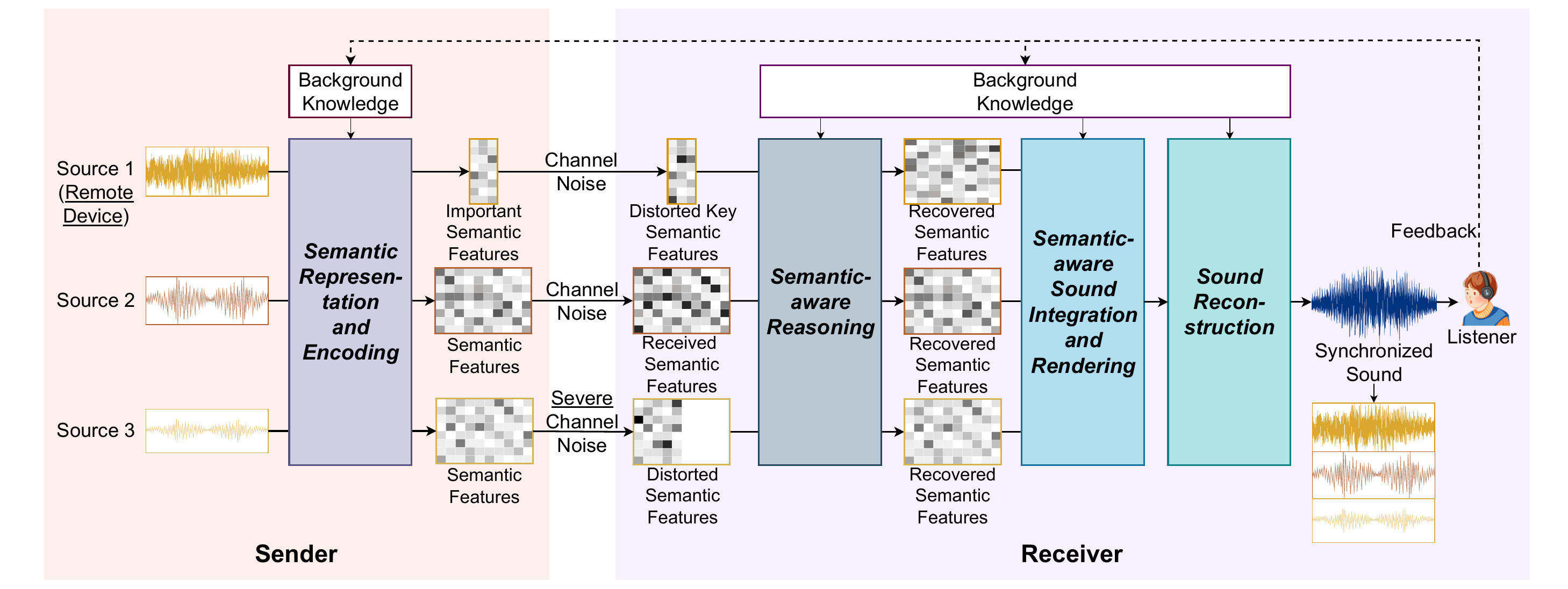}
    \caption{SC for sound quality and timing synchronization.}
    \label{scforsyn}
\end{figure*}

\section{Synchronization challenges in the IoS and How SC Can Help}
In this section, we explore the specific synchronization challenges in the IoS in terms of sound quality, timing and behavior control. We then discuss how SC can help address these challenges. 
\subsection{Synchronization Challenges in the IoS}
The IoS extends the principles of the IoT to the auditory domain. It encompasses a wide range of technologies and techniques for sound processing and transmission while leveraging sophisticated algorithms and large online sound-based repositories. 
As defined in \cite{turchet2023ios}, the IoS represents “the ensemble of sound devices, network infrastructures, protocols, and representations of sound-related information that enable services and applications for the communication of sound-related information in physical and/or digital realms”. 
Sound devices are networked computing devices equipped with sensors and actuators capable of capturing, processing, sharing, or producing sounds and sound-related information. 
The term ``sounds'' is used hereinafter to indicate the union of music, speech, and other audio signals. 
Sound-related information involves perceived and processed data which assist IoS devices to exchange sounds efficiently and precisely. 

\textit{Synchronization} is a critical challenge in the IoS, particularly for time-sensitive applications that require seamless and coherent sound communication across different channels and devices. In such applications, ensuring precise synchronization is essential for delivering a high-quality and immersive auditory experience to users.
The current IoS networks mainly face sync challenges on \textit{sound quality}, \textit{timing} and \textit{behavior control}.

\subsubsection{Sound Quality}
IoS listeners receive sound streams from multiple IoS senders and integrate them. If the quality of certain sound streams is unacceptable, it may divert listeners' attention, detracting from the overall auditory experience.
Imperfect sound integration can result in a fragmented and unsatisfying auditory experience for listeners. Therefore, sound quality synchronization pertains to the fact that the IoS must ensure that all the sound streams delivered to listeners are well-orchestrated and maintain a similar level of fidelity and accuracy. However, the IoS spans a large geographical area, involving multiple wireless links, heterogeneous devices, and varying network conditions. Consequently, sound quality is often compromised due to distortion and interference introduced by poor-quality channels and long-distance transmission.
Moreover, the IoS has a complex sound environment, in which multiple sound sources may coexist, such as speech, music, and background noise. Hence, this complex sound environment further makes it challenging to fulfill the diverse service requirements for various IoS sound types.

\subsubsection{Timing}
IoS listeners expect a seamless coordination of multiple sounds. Even a slight vibration of timing can lead to dissonance, echoes, or a disjointed auditory perception. Thus, precise timing alignment is crucial for maintaining the integrity and coherence of the overall auditory experience. 
To achieve accurate timing, the end-to-end delay consisting of sound processing time and sound transmission time should be considered. The processing latency is determined by the devices' computing capabilities, and thus it falls outside the scope of our discussion. 
Our goal is to reduce sound transmission time and mitigate jitter, as both timely and consistent reception of each sound stream by the receiver are crucial for achieving precise timing synchronization. However, channels with varying transmission rates can introduce gaps in sound streams and unpredictable packet arrival times, which pose key challenges. Consequently, developing low-latency, low-jitter communication schemes that minimize transmission delay and variability is essential to ensure robust timing synchronization.

\subsubsection{Behavior Control}
In a specific sound synchronization task, such as a live concert, each device may exhibit different behavior based on its assigned role. 
For instance, the lead vocalist's microphone would typically be given a higher priority, as the clarity and timing of the vocals are crucial for song delivery. However, during a guitar solo, the IoS devices may need to dynamically shift their focus to prioritize the synchronization of the guitarist's amplifier output.
Therefore, IoS devices should adjust their behavior to accommodate IoS listeners' requirements in a timely manner to deliver a high-quality and immersive auditory experience.

Current IoS networks heavily depend on control messages generated from listeners' feedback to guide device behavior control. However, relying exclusively on listeners feedback can be unreliable, as listeners may require considerable time to provide input. Moreover, this feedback often targets a specific sender, which hinders the system's capacity to learn from diverse experiences. If the same issues occur again, transmitting redundant feedback would lead to a waste of resources. Therefore, introducing a novel behavior control strategy is essential to enhance sound synchronization in the IoS, addressing the limitations of the current user feedback-based approach.

\begin{figure*}[!t]
    \centering \includegraphics[width=0.8\textwidth]{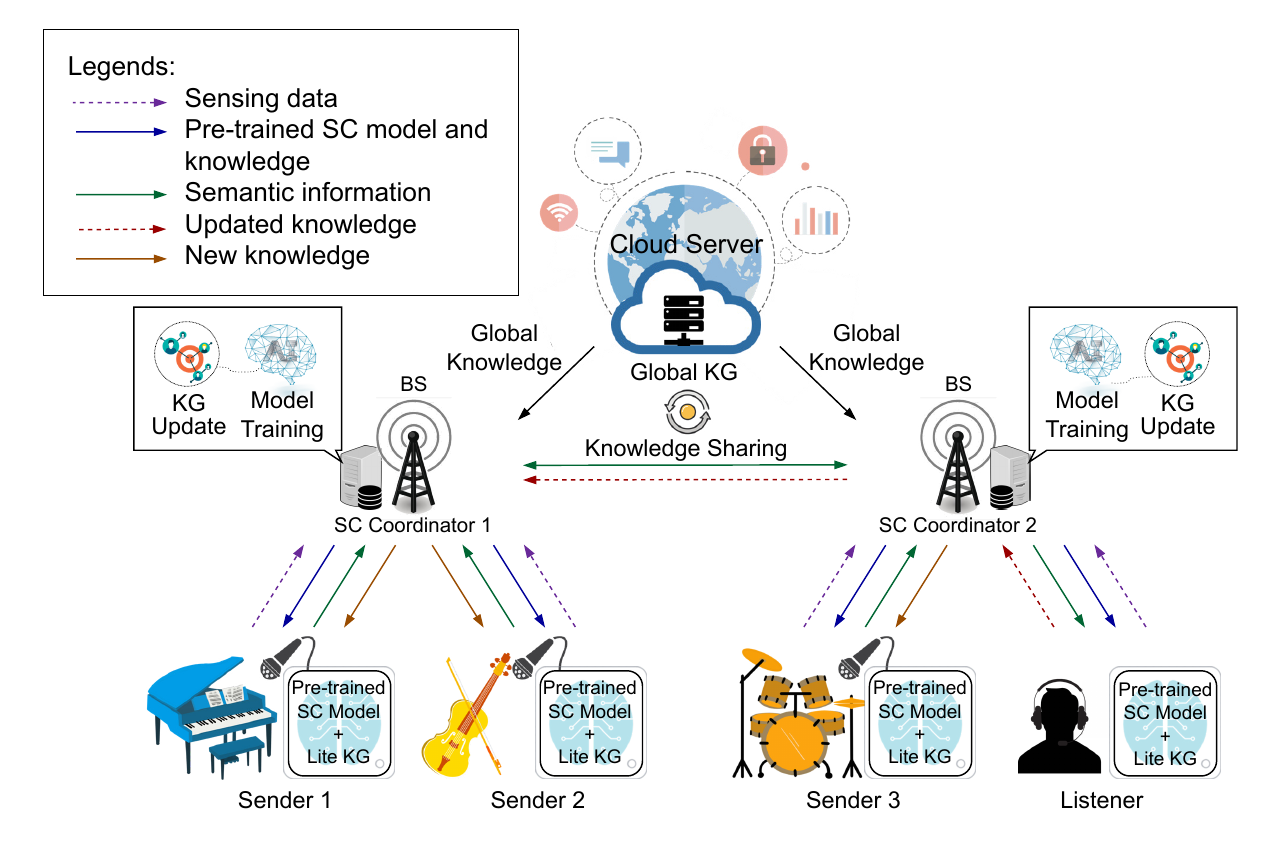}
    \caption{Architecture of multi-user SC enabled IoS networks for synchronization.}
    \label{sync}
\end{figure*}

\subsection{How SC can Help for Synchronization Challenges?}
SC diverges from conventional Shannon communication by integrating human-like ``comprehension" and ``reasoning" into the data encoding, transmission and decoding processes, rather than striving for precise bits duplication \cite{yang2022semantic}. To be concrete, SC systems extract the semantic information from a source based on the background knowledge, transmit it through a physical channel, and reconstruct the source data by reasoning from background knowledge and context. 
Due to its superior understanding and reasoning capabilities, SC could address the sound synchronization challenges in the IoS. Specifically, SC offers significant advantages in sound quality synchronization, timing synchronization, and synchronization behavior control as follow. 


\begin{itemize}
    \item SC enables \textit{sound quality} synchronization by facilitating semantic-aware sound prediction and error correction, particularly for those devices suffering poor channel conditions.
    As shown in Fig.~\ref{scforsyn}, semantic features, which contain contextual and knowledge-based information, are extracted from sound sources. However, these features may become distorted or lost during the transmission process due to severe channel noise and interference. Those distorted or missing features can be inferred from the remaining features by leveraging the contextual relationships and predefined patterns in listeners' knowledge bases. This semantic-aware error resilience approach can improve the quality of sound streams that experience poor channel quality.
    Furthermore, by analyzing semantic information, listeners' device setup, and their preferences, SC systems can adjust the relative levels, panning, and spatial positioning of different sound elements while filtering out irrelevant or conflicting elements. 

    \item SC enables \textit{timing} synchronization by reducing the transmission latency for delayed sounds through prioritizing the transmission of important semantic information.
    As shown in Fig.~\ref{scforsyn}, SC encoders can extract and transmit important semantic features. Once these features are received, the context can be inferred based on their semantic relationships. Even if listeners do not receive the complete message, they can still begin synchronizing the sound streams while inferring the remaining content based on the available semantic information and context. As a result, this approach compensates for transmission delays, which is particularly beneficial for IoS devices suffering from low-transmission-rate channels. 
    \item SC enhances \textit{behavior control} by updating senders' models and sharing knowledge among listeners and senders. New knowledge can be derived from listeners' feedback, sensing data, and history. Edge servers equipped with advanced computing ability can fine-tune SC models using the updated knowledge bases, and distribute them to senders. The updated knowledge bases ensure that senders can react accurately and efficiently if they encounter repetitive problems. However, updating knowledge bases and models will incur overhead and consume time. Hence, edge servers must determine when and how to perform updates based on timing requirements, resource usage, network conditions in the IoS.
\end{itemize}

\begin{figure*}[!t]
    \centering
    \includegraphics[width=1.01\textwidth]{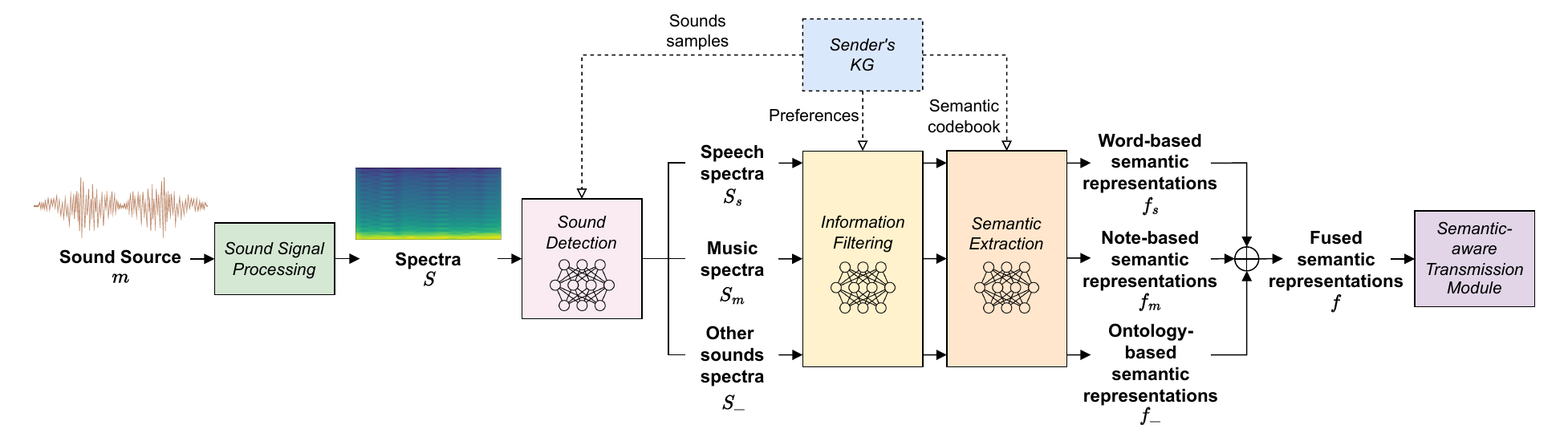}
    \caption{Semantic encoding in IoS-centric SC systems.}
    \label{encoder}
\end{figure*}

\section{IoS-Centric Semantic Communication Framework}
As shown in Fig.~\ref{sync}, an IoS-centric network consists of sound devices (senders and listeners), SC coordinators, BSs and cloud servers.
Sound devices are equipped with sensors or actuators that enable them to capture, process, share, and produce sounds and sound-related information.
SC coordinators, which can be edge servers deployed on BSs, play a crucial role in pre-training SC models, sharing knowledge among sound devices and cloud servers, as well as updating knowledge bases to control sound devices behavior.
BSs are responsible for sounds and sound-related information exchanges among sound devices, SC coordinators, and cloud servers.
Cloud servers facilitate the global interconnection of SC coordinators and provide global sound repositories.

To achieve effective sound synchronization, a tight collaboration among these entities is needed. Initially, sound devices monitor their surroundings and capture environmental sounds to gain valuable insights into the acoustic context. For instance, if a microphone consistently detects piano sounds, it will prioritize capturing and processing these specific sounds to improve its performance in synchronization.
The sensing data collected from sound devices is a kind of private knowledge which will be uploaded to SC coordinators. 
SC coordinators collect these private knowledge, download global knowledge from cloud servers, and then merge them together into a structured form, i.e., a knowledge graph (KG) \cite{Ji_2022}. 
Subsequently, SC coordinators use KGs to pre-train SC models, and then distribute the pre-trained SC model and a local KG to each sound device. Due to the limited storage and computing abilities of sound devices, local KGs are oriented towards device behavior and the preferences of the users while filtering out irrelevant and unnecessary global knowledge \cite{liang2023generative}. Both SC coding models and KGs are updated periodically, which will not generate an additional delay for particular sound transmissions. 

Given the SC coding models and KGs, senders extract semantic features from the source and encode it into symbols for wireless transmission to listeners. 
Next, listeners decode the received symbols and reconstruct the original sound signals from semantic features based on their KGs. In scenarios involving multiple sound sources, listeners also integrate and refine the received sounds, aiming to enhance the quality of synchronization based on the preferences of the users. For instance, if users have a preference on piano sounds, the devices will prioritize and amplify the piano sounds during the sound synchronization process. 

Subsequently, SC coordinators analyze listeners’ feedback, sensing data, and history to extract valuable insights and update their KGs by integrating new knowledge.
By leveraging the updated KGs, SC coordinators fine-tune the SC models. This fine-tuning process involves adjusting the model parameters and architecture to better capture the semantic relationships and optimize the behavior performance of sound devices. The fine-tuned SC models are then distributed to the sound devices along with the relevant local KGs, tailored to their specific behavior and requirements.

\section{Transceiver Design in IoS-centric SC Networks}
Building upon the IoS-centric SC networks framework, we present a transceiver design, including semantic encoding, semantic importance-aware transmission, and semantic decoding. In particular, we present a symbolic semantic encoding module designed to represent sounds accurately and efficiently. This robust semantic encoding facilitates high-fidelity sound reconstruction, contributing significantly to sound quality synchronization. Concurrently, our semantic decoding process maintains sound quality by reasoning through distorted or missing sound elements and rendering sounds based on their semantic information. To achieve timing synchronization, we introduce a semantic importance-aware transmission system. This system prioritizes the transmission of important semantic information while filtering out irrelevant data, thereby reducing overall transmission latency.

\begin{figure*}[!t]
    \centering
    \includegraphics[width=1.0\textwidth]{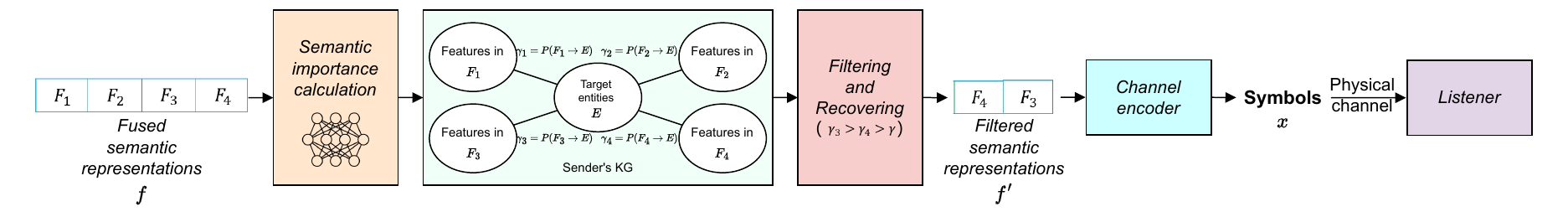}
    \caption{Semantic importance-aware transmission in IoS-centric SC systems.}
    \label{siat}
\end{figure*}

\begin{figure*}[!t]
    \centering
    \includegraphics[width=1.0\textwidth]{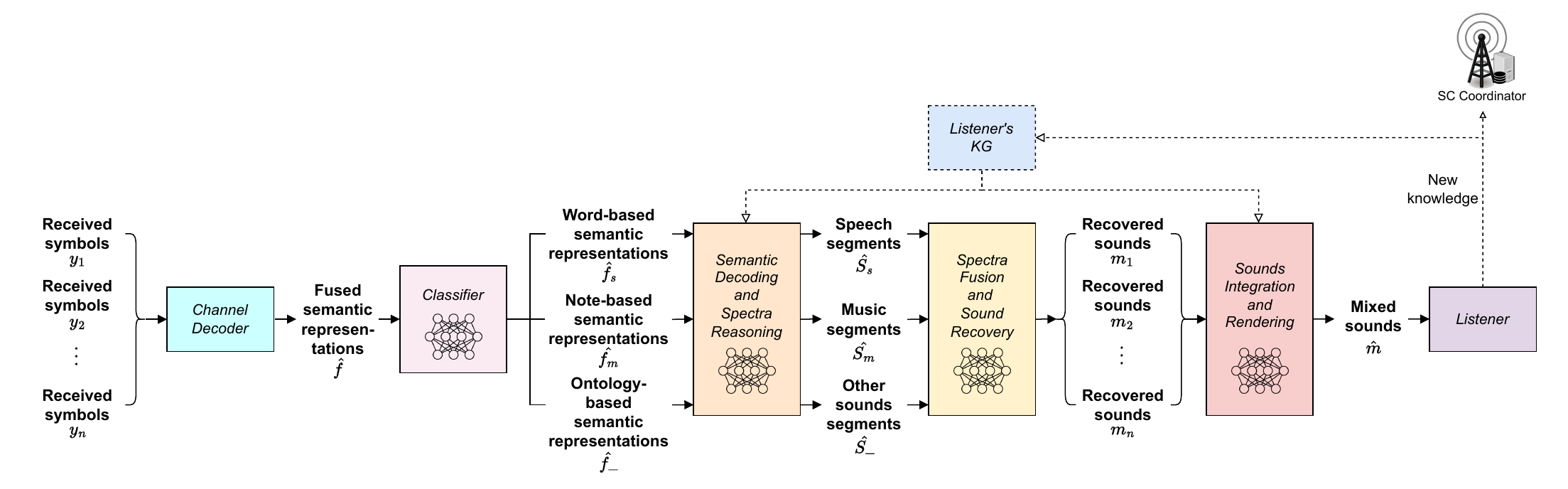}
    \caption{Channel decoding and semantic decoding in IoS-centric SC systems.}
    \label{decoder}
\end{figure*}

\subsection{Semantic Encoding}
Contextual and knowledge-assisted sound prediction, coupled with semantic-aware sound rendering plays a crucial role in achieving sound quality synchronization. To effectively implement these functions, it is essential to accurately represent sound in a logical manner leveraging contextual relationships and external knowledge. Although there are various types of sounds with diverse nature in IoS networks, such as speech, music, and background sounds, unified semantic representation methods can be developed for representing them. Neurosymbolic AI techniques, which stem from neural networks and symbolic AI, offer feasible solutions by combining reasoning with complex representations of knowledge, such as KGs and ontologies \cite{CKT,saad2024artificial}.

However, there are slight differences in the symbolic units among speech, music, and background sounds. Speech can be transcribed into a sequence of word tokens, where each token represents a discrete unit of meaning derived from a predefined vocabulary or alphabet. Similarly, borrowing ideas from \cite{guo2023domain}, music can be symbolized by notes, which include pitch, duration, and onset information as fundamental music tokens. The semantic representations for music are then derived from these tokens.
For sounds with less informative content, such as wildlife or machine noises, semantic representations can be extracted based on key attributes including the sound's source, environmental context, temporal characteristics, and other relevant acoustic features. 

Before encoding different types of sounds, it is necessary to detect and extract them from a source.
To achieve this, a few traditional sound processing steps should be conducted. As shown in Fig.~\ref{encoder}, a sound signal sequence is first converted into a spectrogram \cite{kingsbury1998robust}, which is a visual representation of the frequencies present in the sound signal and how they change over time. 
Next, acoustic features are extracted from the spectrogram. These features may include spectral characteristics, temporal patterns, and other relevant information that can help identify and distinguish different types of sounds \cite{mesaros2021sound}. By comparing the extracted features with predefined sound templates, the category of each sound segment can be detected. 

Subsequently, a filtering process is applied to remove sounds that are not pertinent to the sender's intended behavior. 
The filtered sound segments then undergo semantic representation and encoding. By querying acoustic features in the sender' KG and searching for their corresponding semantic representations, the speech, music and other segments are transformed into word-based, note-based and ontology-based semantic representations respectively. 
Finally, all the semantic representations extracted from the different sound types are fused together. This fusion process combines the semantic information from speech, music, and background sounds into a unified representation. The fused semantic representations are then fed into the semantic importance-aware transmission module.

\subsection{Semantic Importance-aware Transmission Module}
To enhance timing synchronization, particularly for low-bandwidth scenarios, a semantic importance-aware transmission module can be implemented. This module adapts semantic representations based on semantic importance of source data and channel conditions. For instance, for low-rate links, there may be no need to transmit background sounds that a listener may be not interested in.
In particular, in a fixed-length semantic representation, each token is reordered based on its semantic importance. A token that is more closely related to the listener's preferences is considered more important. Unlike classical priority-based queuing, which relies on bit-oriented rules for packet prioritization, the semantic importance-aware transmission approach prioritizes sound segments based on their perceived relevance to the listener.

As shown in Fig.~\ref{siat}, the semantic importance of a token is quantified by calculating the relevance between this token and the listener's preferences. 
The relevance calculation follows the True Logic where the relation in a triple (head, relation, tail) indicates the truth value that the tail is true if the head is true (head $\to$ tail) \cite{choi2022unified}. The semantic representations of a token and target entities related to listeners' preferences have been located in the sender's KG during semantic encoding and model training. To reduce querying delay, advanced graph traversal algorithms are employed to efficiently search for target entities from tokens along the shortest paths within this KG. The probability calculated along the shortest path represents the semantic relevance between the token and the listener's preferences, indicating the token's importance. Meanwhile, the less important tokens can be selectively filtered out if messages are transmitted through low-transmission-rate channels. When the listener receives important semantic data, they can begin sound reconstruction promptly to enable faster sound processing and playback.

\subsection{Semantic Decoding}
To enhance the sound quality of synchronization, decoders exploit semantic-aware reasoning techniques that leverage contextual information and knowledge to intelligently reconstruct and interpolate missing or corrupted sound segments. As shown in Fig.~\ref{decoder}, the received semantic representations are recovered and decoded into word-based, note-based, and ontology-based semantic representations. Subsequently, these semantic representations are transformed back into their corresponding acoustic features in a spectrogram based on the listener's KG.

However, the recovered sound quality may be poor due to adverse communication conditions, such as noise, interference, or packet loss. Inspired by \cite{han2022semantic}, the plausible hypotheses for the missing or distorted sound segments can be predicted from context and predefined relationships between entities in the listener's KG. 
Moreover, we note that this KG-based approach may not be suitable for handling entirely new or unseen information that cannot be found in the existing KG. In such cases, the system relies solely on the contextual relationships to reason about the missing content. 

Furthermore, to achieve sound quality synchronization, the reconstructed sound segments undergo precise adjustment and rendering during the sound integration process. This procedure harmonizes the acoustic resonance and filters out discordant elements, thereby enhancing the listener's auditory experience. First, the reconstructed sound segments are they are seamlessly fused into a unified spectrogram. Then, the semantic decoder analyzes the semantic content within the sounds, taking into account factors such as the listener's device setup and personal preferences. Based on this analysis, the semantic decoder intelligently adjusts the relative levels, panning, and spatial positioning of different sound elements, creating a balanced and immersive soundscape \cite{dai2022binaural}. It prioritizes the most semantically relevant sound elements based on listeners' preferences, while filtering out irrelevant or conflicting elements that may detract from the desired listening experience. 
Finally, the spectrogram is converted back into sound waves using techniques like inverse Fourier transform or wavelet synthesis. These sound waves can then be played through the listener's audio output devices.

\section{Open Issues and Discussion}
While the proposed IoS-centric SC networks shows numerous advantages, it also presents several unavoidable and complex challenges that should be addressed to fully unlock its potential.

\textbf{Mathematical SC Models:}
A significant challenge arises in developing rigorous mathematical models for integrating KGs and SC in the context of the IoS. Currently, it is not clear how KGs should explicitly affect semantic encoding and decoding, particularly in terms of the impact on semantic entropy and semantic ambiguity. For IoS applications, this uncertainty extends to how KGs can optimally represent temporal sound sequences, acoustic environments, and cross-modal relationships between sounds and other sensory data. Understanding this impact is crucial for theoretically formulating the sound quality synchronization problem in our IoS-centric SC networks. Some mathematical tools such as logic probability, category theory, and neurosymbolic models could be promising here. 

\textbf{Adaptive Resource Allocation Strategies:}
Designing resource allocation strategies for timing synchronization in IoS networks presents a significant challenge. Efficient strategies can prioritize the allocation of resources to transmit important semantic data based on channel conditions and timing requirements, ensuring accurate timing synchronization across multiple sound streams. This is particularly important in IoS-centric SC systems in which network conditions, semantic importance, and user preferences are often dynamic and unpredictable. Consequently, novel resource allocation strategies are needed to optimize both timing and quality synchronization in IoS networks, taking into account SC features.

\textbf{Cross-layer Protocol Design:}
To enhance IoS applications, particularly in terms of sound quality synchronization performance, the joint design of application layer protocols and wireless data transmission protocols is important yet challenging. Cloud-based applications in IoS-centric SC networks often rely on diverse global sound repositories containing data in various formats and types. This heterogeneity between global repositories and local knowledge can lead to semantic ambiguity between the application layer and the wireless transmission layer, potentially degrading the accuracy and efficiency of sound quality synchronization.
The semantic inconsistencies across layers can result in misinterpretation of acoustic features, tonal characteristics, and contextual cues essential for precise sound quality synchronization.
Hence, a novel cross-layer protocol incorporating a unified semantic language is required to ensure that semantic data can be effectively and accurately represented, transmitted, and interpreted across the different layers of IoS networks. 

\section{Conclusions}
In this article, we have delved into integrating SC into the IoS, with a particular emphasis on addressing the synchronization challenges in the IoS. Besides offering visions at a conceptual level, we have proposed a new framework of IoS-centric SC networks and illustrated the transceiver design to address the synchronization challenges. Finally, we have discussed several open issues in terms of mathematical models, resource allocation strategies, and cross-layer communication protocol design for the IoS. We hope this research serves as a pioneer in exploring SC for synchronization and the IoS, paving the way for advanced semantic-driven wireless sound delivery.

\bibliographystyle{IEEEtran}
    \bibliography{ref}
\end{document}